# Evidence of Charge-Phonon coupling in Van der Waals materials Ni$_{1-x}$Zn$_x$PS$_3$

Nashra Pistawala[1§], Ankit Kumar[1§], Devesh Negi[3§], Dibyata Rout[1], Luminita Harnagea[2], Surajit Saha[3] and Surjeet Singh[1 *]

[1]Department of Physics, Indian Institute of Science Education and Research, Pune-411008
[2]I-HUB Quantum Technology Foundation, Indian Institute of Science Education and Research, Pune-411008
[3]Department of Physics, Indian Institute of Science Education and Research, Bhopal-462066

§ Equal contributions
* surjeet.singh@iiserpune.ac.in

**ABSTRACT**

NiPS$_3$ is a Van der Waals antiferromagnet that has been found to display spin-charge and spin-phonon coupling in its antiferromagnetically ordered state below $T_N$ = 155 K. Here, we study high-quality crystals of site-diluted Ni$_{1-x}$Zn$_x$PS$_3$ (0 < x < 0.2) using temperature dependent specific heat and Raman spectroscopy probes. The site-dilution suppresses the antiferromagnetic ordering in accordance with the mean-field prediction. In NiPS$_3$, we show that the phonon mode P$_2$ (176 cm$^{-1}$) associated with Ni vibrations show a distinct asymmetry due to the Fano resonance, which persists only in the paramagnetic phase, disappearing below $T_N$ = 155 K. This was further supported by temperature dependent Raman data on an 8% Zn-doped crystal ($T_N$ = 135 K) where Fano resonance similarly van in the magnetically ordered phase. This is contrary to the behavior of the Raman mode P$_9$ (570 cm$^{-1}$), which shows a Fano resonance at low temperatures below $T_N$ due to its coupling with the two-magnon continuum.

We show that the Fano resonance of P$_2$ arises from its coupling with an electronic continuum that weakens considerably upon cooling to low temperatures. In the doped crystals, the Fano coupling is found to enhance with Zn-doping. These observations suggest the presence of strong electron-phonon coupling in the paramagnetic phase of NiPS$_3$ due to charge density fluctuations associated with the negative charge transfer state of Ni.

**INTRODUCTION**

Two-dimensional (2D) Van der Waals materials have attracted enormous attention in recent years. While most two-dimensional Van der Waals materials are non-magnetic, the inclusion of magnetic degrees of freedom results in rich physics with intriguing magnetic phases unique to the 2D world. Recent breakthroughs in detecting magnetic long-range ordering in 2D magnetic materials down to atomically thin limits, as demonstrated in FePS$_3$ [1], CrI$_3$ [2] Cr$_2$Ge$_2$Te$_6$ [3] have opened up new and remarkable possibilities for probing and manipulating the magnetic properties of these materials.

The Transition metal chalcogenophosphates MPX$_3$ where M = {Ni, Fe, Mn, Co, or Zn}, and X = {S or Se} are interesting magnetic Van der Waals materials with magnetic M atoms residing on a honeycomb lattice [4–7]. While all known MPX$_3$ compounds undergo some form of long-range antiferromagnetic ordering at low temperatures, the exact nature of the magnetic ground state depends on the choice of the transition metal ion, ranging from an Ising-type antiferromagnetic in FePS$_3$ [8], XY or XXZ-type antiferromagnetic in NiPS$_3$ [9], and nearly Heisenberg-type antiferromagnetic in MnPS$_3$ [10]. The Néel temperature (T$_N$) varies with the transition metal ion as: 78 K (MnPS$_3$) [11], 118 K (FePS$_3$) [12], 122 K (CoPS$_3$) [13], and NiPS$_3$ (155 K) [9]. The choice of transition metal ion also gives rise to a tunable electronic structure with widely ranging band gaps varying from 1.5 eV (FePS$_3$) to 3.5 eV (ZnPS$_3$) [14]. The recent studies unveiled emergence of interesting phenomena, including insulator-to-metal transition in MnPS$_3$, MnPSe$_3$, FePS$_3$, NiPS$_3$ [16–18], and V$_{0.9}$PS$_3$ [19], and superconductivity in FePSe$_3$ [20].

Here, we focus on NiPS$_3$. Analogous to the other members of MPX$_3$ family, NiPS$_3$ has a layered crystal structure with weak Van der Waals bonding between the adjacent layers stacked along the $c^*$ direction, normal to the $ab$-plane as show in Fig. 1(a). In each layer, the Ni$^{2+}$ ions form a honeycomb network. The P and S atoms bond covalently to form [P$_2$S$_6$]$^{4-}$ anionic units. These units sit at the center of the hexagon formed by Ni with their P-P bond aligned perpendicular to the honeycomb plane of Ni as shown in Fig. 1b. This way, each Ni atom is surrounded by six S atoms in a trigonally distorted octahedral coordination (Fig. 1b). In the ordered phase below $T_N$ = 155 K [9], the Ni spins, confined to the $ab$-plane with a slight out of plane canting, form a zigzag antiferromagnetic pattern where ferromagnetic zigzag chains running parallel to the $a$-axis couple antiferromagnetically along the $b$-axis so that the spin direction rotate by 180° between the alternating chains [15], as shown in Fig. 1a.

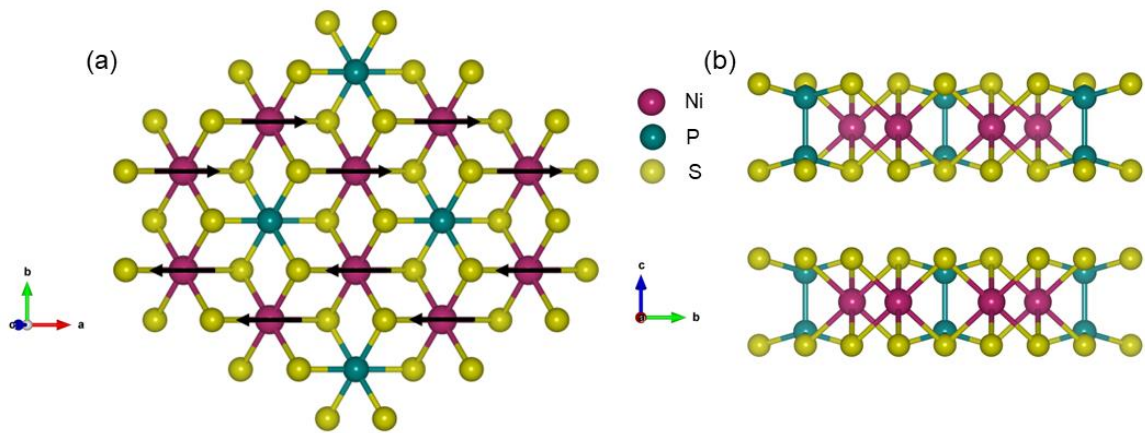

**Fig. 1**. Schematic representation of crystal structure. The pink, green, and yellow balls represent Ni, P, and S, respectively. (a) Structure as viewed along the $c^*$-axis perpendicular to the $ab$-plane showing the Van der Waals layer with Ni atoms forming a distorted honeycomb lattice. The black arrows indicate the alignment of Ni spins. (b) The structure as viewed along the $a$-axis showing the layered structure of NiPS$_3$. Within each layer, the P-P dimer bonds with six S atoms to form [P$_2$S$_6$]$^{4-}$ anionic unit.

Recently, it has been shown that NiPS$_3$ is a 'self-doped' negative charge transfer (NCT) insulator where the *p* orbitals of S display a strong hole-like character due to a negative charge transfer energy. The Ni *d* orbitals accordingly form a linear combination of d$^8$, d$^9\underline{L}$, and d$^{10}\underline{L}^2$ configurations, where $\underline{L}$ refers to a ligand (S) hole. The NCT state in NiPS$_3$ has been supported by the x-ray absorption and photoemission experiments [16, 17].

Intriguingly, below T$_N$ the photoluminescence (PL) spectra of NiPS$_3$ exhibits a sharp, spin-orbit entangled excitonic peak around 1.47 eV [18]. This peak becomes exceedingly sharp (FWFM ≈ 0.4 meV) upon cooling below a temperature T$_{coh}$ ≈ 50 K [19]. Since the bandgap of NiPS$_3$ (≈1.6 eV) [14] is larger than the energy of the PL peak, the resolution limited FWHM below T$_{coh}$ is unusual and has been attributed to the coherent excitonic emission due to Zhang-Rice (ZR) triplet to singlet transition [19]. The PL peak is also shown to disappear in the monolayer limit, which coincides with the vanishing of the long-range antiferromagnetic ordering for a monolayer NiPS$_3$. This observation has been used to suggest a correlation between the magnetic ordering and the appearance of the PL peak [18].

Recently, Raman spectroscopy has emerged as a powerful tool to investigate 2D antiferromagnetic Van der Waals materials, particularly in the few-layers and monolayer limit where the conventional techniques for the bulk samples are not sensitive enough to pick the signature of magnetic long-range ordering due to small sample mass [20–23]. In NiPS$_3$, the temperature and polarization-dependent Raman spectroscopy, down to the monolayer limit, revealed some very intriguing features [24]. Firstly, a broad peak is seen in the Raman spectra centered around 550 cm$^{-1}$ due to two-magnon scattering. This peak is particularly pronounced in the ordered state at low temperatures (T ≲ T$_N$). Secondly, one of the Raman active modes (P$_9$), which lies in the same energy range as the two-magnon peak, exhibits a prominent Breit−Wigner−Fano (BWF) lineshape upon cooling below T$_N$ due quantum interference between the P$_9$ phonon mode and the two-magnon continuum [24]. Equally intriguingly, another Raman mode (P$_2$) at 176 cm$^{-1}$ splits in the parallel- and cross-polarization geometries when the sample is cooled below T$_N$. However, above T$_N$ the splitting disappears. P$_2$ is a composite mode comprising two degenerate phonon modes of symmetry A$_g$ and B$_g$. Both these modes involve vibration of Ni ions with the difference that in A$_g$ the displacement is perpendicular to the ordered moment and in B$_g$ it is parallel, this distinction disappears in the paramagnetic phase. Using these two features (the splitting of P$_2$ phonon mode and the Fano resonance of the P$_9$ mode, both appearing below T$_N$), it was shown that antiferromagnetic ordering in NiPS$_3$ persists down to the two layers with an almost unsuppressed T$_N$ before vanishing completely in the monolayer limit. However, persistent spin fluctuations were seen even in the monolayer NiPS$_3$ as expected in the XY spin model. These observation, along with the optical conductivity study in Ref. [16], suggest the presence of strong spin-phonon and spin-charge coupling in NiPS$_3$.

In this work, we present a systematic study of site-diluted Ni$_{1-x}$Zn$_x$PS$_3$ ($0 \leq x \leq 0.2$). The investigations are carried out on high-quality single crystals using temperature dependent specific heat and Raman spectroscopy probes. More specifically, we carefully probe the behavior of mode P$_2$ near 176 cm$^{-1}$ as a function of temperature and Zn-doping (x). We show that P$_2$ exhibits a Fano-like asymmetry in the paramagnetic phase above T$_N$(x) due to its coupling with a broad electronic continuum. This continuum disappears in the magnetically ordered phase making P2 symmetric (Lorentzian) at low temperatures. We show that the value of coupling coefficient increases with Zn doping. We therefore propose the existence of strong charge-phonon coupling in the paramagnetic phase of NiPS$_3$.

I. **EXPERIMENTAL METHODS**

A. **Crystal growth**

High-quality single crystals of $Ni_{1-x}Zn_xPS_3$ were grown using the physical vapor transport technique. The starting materials were high purity elemental powders Nickel (Sigma Aldrich 99.99% trace metal basis), Zinc (Sigma Aldrich 99%), Phosphorous (Sigma Aldrich 99.99%), and Sulphur (Sigma Aldrich 99.98% trace metal basis). The growth experiments were carried out in sealed silica ampoules. All the room-temperature processes were carried out in an argon-filled glove box ($O_2$ and $H_2O$ contents less than 0.1 ppm). More details concerning the crystal growth parameters can be seen in Ref. [25]. The shiny black crystals thus obtained exhibit hexagonal shape with lateral dimensions ranging up to 1 cm and thickness ≤ 1mm, as shown in the inset of Fig. 2. The crystals showed a tendency to exfoliate rather easily.

**B.     Experimental techniques**

The morphology and chemical composition of the crystals was verified using the field-effect scanning electron microscope (Zeiss Ultra Plus) equipped with Energy Dispersive X-ray (EDX) (Oxford Instruments) analysis, respectively. The phase purity of the grown crystals was checked using powder x-ray diffraction (Bruker D8 Advance diffractometer). For obtaining the powders, small crystal specimens were ground in an agate mortar-pestle. For an accurate determination of the lattice parameters, high-purity silicon was added as an internal standard. The quality of grown crystals was assessed using high-resolution Transmission Electron Microscopy (HRTEM). The high-resolution TEM micrographs and corresponding SAED patterns were collected using JEOL JEM 2200FS 200 keV TEM. For TEM sample preparation, the crystal specimens were sonicated in a cyclohexane bath and a small portion of the dispersed solution was drop casted on a carbon-coated copper grid that was dried under vacuum for 3-4 h. The analysis of HRTEM images and SAED patterns was done using GSM-3 package. Raman spectra were collected on thin single crystal specimens in the backscattering configuration with incident Laser light perpendicular to the crystalline $ab$-plane. The Horiba Jobin-Yvon

LabRAM HR spectrometer equipped with liquid nitrogen cooled Charge-Coupled Detector (CCD) and laser of 633 nm as a source of excitation was used. The excitation was maintained at 25% of the maximum power. The data acquisition time for each spectrum was 30 s with 15 iterations to get better resolution and intensity of the Raman modes. These parameters were kept constant across all samples for better comparison. The Linkam stage was used for the temperature-dependent measurements. The specific-heat was measured using a Physical Property Measurement System (PPMS), Quantum Design, USA. A crystal measuring 2 mm by 2 mm, cut from a larger crystal specimen, was mounted on the heat capacity sample holder using a low-temperature APIZON N grease. The heat capacity of the N grease was measured as addenda correction prior to loading the crystal specimen.

## III. RESULTS AND DISCUSSION

### A. Structural analysis

The layered morphology of the grown crystals is clearly visible in the in-lens images shown in Fig. 2(a, b). The EDX composition was obtained by taking area scans at multiple regions on the crystal surface. The chemical homogeneity was verified using the elemental chemical mapping (see Fig. A1, Appendix). The nominal and EDX compositions for our crystals is shown in Table I (Appendix). The actual Zn concentration is found to be somewhat less than the nominal composition. This is particularly evident in samples with higher Zn doping. All through the manuscript, the EDX composition has been used both for analysis and discussion. The presence of preferred orientation is clearly reflected in the powder pattern, where strong reflection due to preferred (00l) orientation can be seen (see, Fig. A2, Appendix). This is due to the platelet-like morphology of the grown crystals. The variation of lattice parameters with Zn doping concentration is shown in Fig. A2(b), Appendix (see also Table I in Appendix). The lattice parameters show a monotonic increase consistent with the larger ionic radius of $Zn^{2+}$

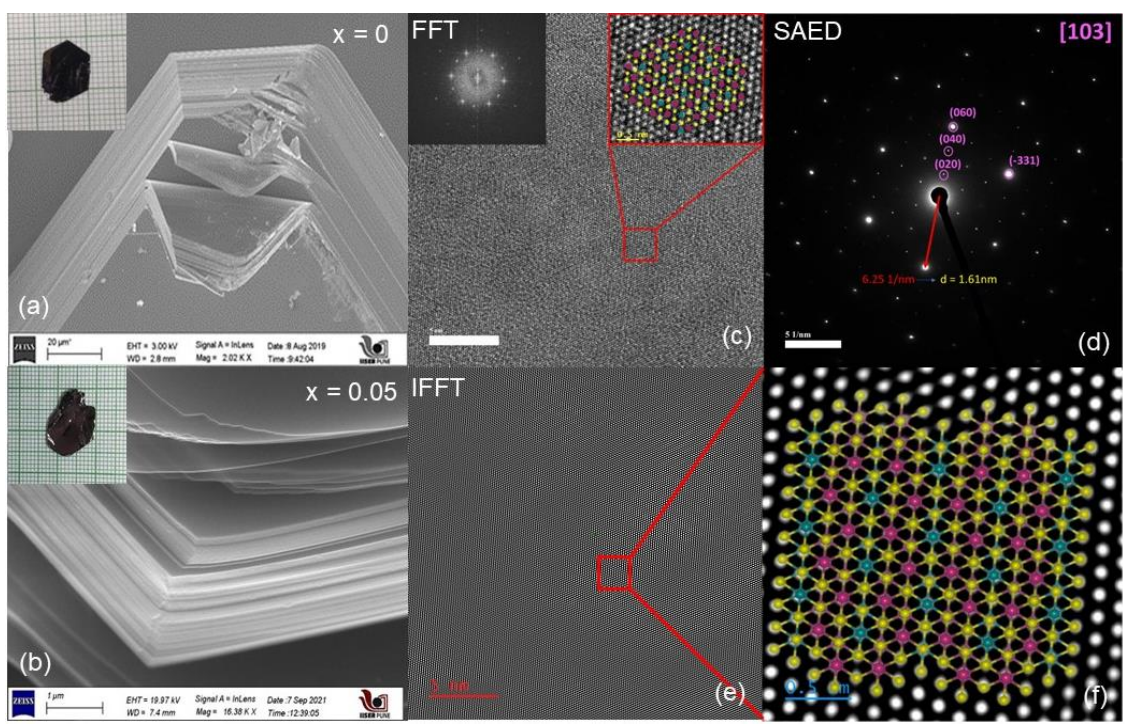

**Fig 2.** (a, b) A few representative FESEM images for $Ni_{1-x}Zn_xPS_3$. The EDX composition (x) of Zn is given at the top right corner. Inset shows representative images of the respective crystals. (c) High-Resolution TEM micrograph for $NiPS_3$. Insets (left) show the corresponding FFT pattern; the arrangement of Ni (Pink), P (Green), and S (yellow) atoms (right) is also shown in the right inset where a magnified view is shown. (d) SAED pattern for $NiPS_3$ (e) The IFFT pattern of the HRTEM image in (a) (f) IFFT pattern with Ni, P, and S atoms are also shown perpendicular to the [1 0 3] direction.

(0.74) compared to $Ni^{2+}$ (0.69). Fig. 2(c) shows a representative high-resolution TEM image of a $NiPS_3$ crystal. The micrograph confirms the hexagonal arrangement of elements mapped in the inset, and in the corresponding FFT pattern. Fig. 2d shows the corresponding SAED pattern perpendicular to the *ab*-plane (along $c^*$). The sharp diffraction spots confirm the high quality of our single crystals. Fig. 2e shows the IFFT of the micrograph in Fig. 2c. To index the SAED pattern, we used the general formula for the angle between the two (h k l) indices in a monoclinic system. This is confirmed by measuring the angle (60°) between the diffraction spots (0 6 0) and (-3 3 1). As discussed in earlier reports on $FePS_3$ [26] and $Fe_{2-x}Co_xP_2S_6$ [27], these materials exhibit a pseudo threefold rotational symmetry by the formation of rotational twin structure along the [1 0 3] direction in the monoclinic system, a similar three-fold symmetry is evident from the SAED pattern in our crystals.

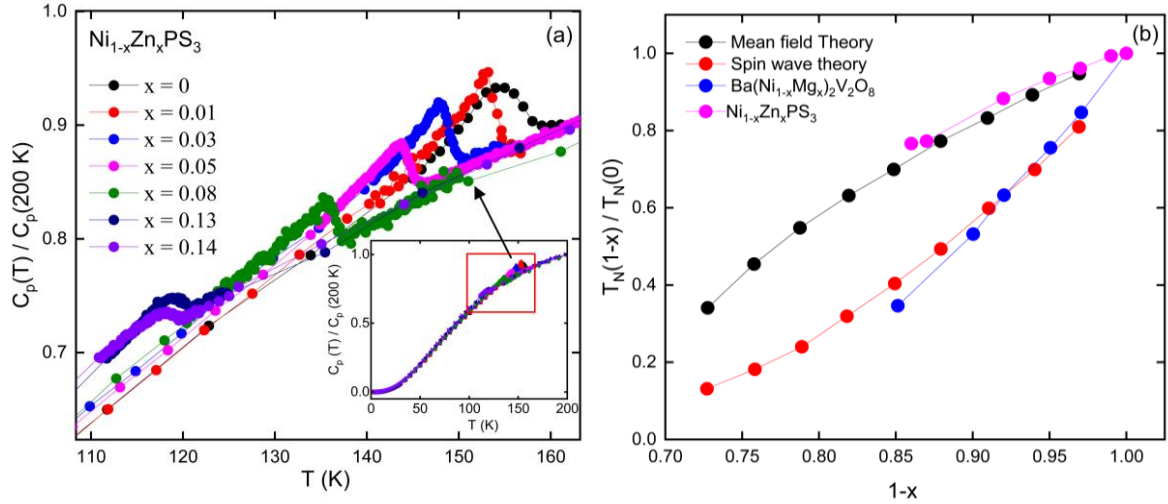

**Fig. 3.** (a) The temperature variation of specific heat ($C_p$) of $Ni_{1-x}Zn_xPS_3$ crystals around $T_N(x)$. For each sample the $C_p$ normalized with respect to its value at 200 K is plotted. Inset shows the specific heat in the range 2 K to 200 K. (b) The plot of $T_N(1-x)/T_N(0)$ against x. The $T_N$ suppression rate from the mean-field theory and spin-wave theory are also shown [29]. The variation of $T_N(1-x)/T_N(0)$ in the honeycomb lattice compound $Ba(Ni_{1-x}Mg_x)_2V_2O_8$ is also shown [30].

### B. Specific Heat

The specific heat ($C_P$) of our $Ni_{1-x}Zn_xPS_3$ crystals is shown in Fig. 3a. In $NiPS_3$, the specific heat exhibits a sharp anomaly near 155 K due to the onset of antiferromagnetic ordering of Ni spins, in agreement with the previous reports [28]. Due to non-magnetic Zn doping at the Ni site the antiferromagnetic ordering temperature decreases as shown in Fig. 3b. The monotonic decrease of $T_N$ due to non-magnetic Zn at the Ni site is consistent with the dilution effect with the rate of decrease of $T_N$ in very good agreement with the mean-field theory for spin 1 on a honeycomb lattice [29]. The departure from the quantum linear spin wave theory, which predicts a much steeper $T_N$ suppression rate, is surprising and require further attention. For comparison, the non-magnetic site-dilution due to Mg in the honeycomb lattice system Ba(Ni$_{1-x}$Mg$_x$)$_2$V$_2$O$_8$ from Ref. [30] is also shown. Interestingly, in this case the $T_N$ suppression follows the quantum linear spin wave theory. The departure in NiPS$_3$ can possibly arise from the fact that the Ni atoms form a distorted hexagonal lattice.

### C. Raman

As shown by Kim et al., the irreducible representations for zone-center phonon modes in NiPS$_3$ are $\Gamma = 8A_g + 7B_g + 6A_u + 9B_u$ [31]. Fig. 4a shows the room-temperature unpolarized Raman spectra of our NiPS$_3$ crystal. All the observed phonon modes are in good agreement with those reported in previous studies [31–33]. We have labelled these modes as $P_1$, $P_2$, $P_3$, etc., in order of increasing wavenumber in accordance with Ref. [24], and using the same peak labels as used there. The Raman modes at higher frequencies (above $\approx$ 200 cm$^{-1}$) are attributed to the intramolecular vibrations of the (P$_2$S$_6$)$^{4-}$ bipyramidal structure, whereas the low-frequency modes $P_1$ and $P_2$ are due to vibrations of Ni [32,33]. As shown in Ref. [24], below $T_N$, the mode $P_2$ near 176 cm$^{-1}$ appears at slightly different positions under different polarization configurations. This is related to the fact that $P_2$ is a composite mode, consisting of phonons with symmetries $A_g$ and $B_g$. Both these modes involve vibration of Ni ions, but in the $A_g$ mode the displacement of Ni is perpendicular to the direction of the ordered moment whereas in $B_g$ it is parallel. This distinction, however, disappears in the paramagnetic phase and the peak $P_2$ appears at the same position in either polarization configuration [24]. The splitting $\Delta P = P_2^{\parallel} - P_2^{\perp}$ has, therefore, been treated as an order-parameter which is non-zero only below $T_N$ and grows analogously to the bulk magnetization for $T < T_N$. Hence, $\Delta P$ has been used as a probe to show that the magnetic ordering in NiPS$_3$ disappears in the monolayer limit [24]. Besides the point that ΔP → 0 for T > $T_N$, for the reasons discussed above, nothing else has been said or known about this mode in the paramagnetic phase in previous literature. Upon careful examination of our 300 K Raman spectra, we found that $P_2$ exhibits a noticeable asymmetry. Attempts to fit $P_2$ using two Lorentzian lineshapes did not yield satisfactory result. This is highlighted in Fig. 4b where fittings with various other lineshapes, including Lorentzian, Voigt, Gaussian, and Fano are also shown. The lineshape that best describes the $P_2$ mode at 300 K turns out to be the Breit-Wigner-Fano (BWF) lineshape [34,35], which is given by the expresssion:

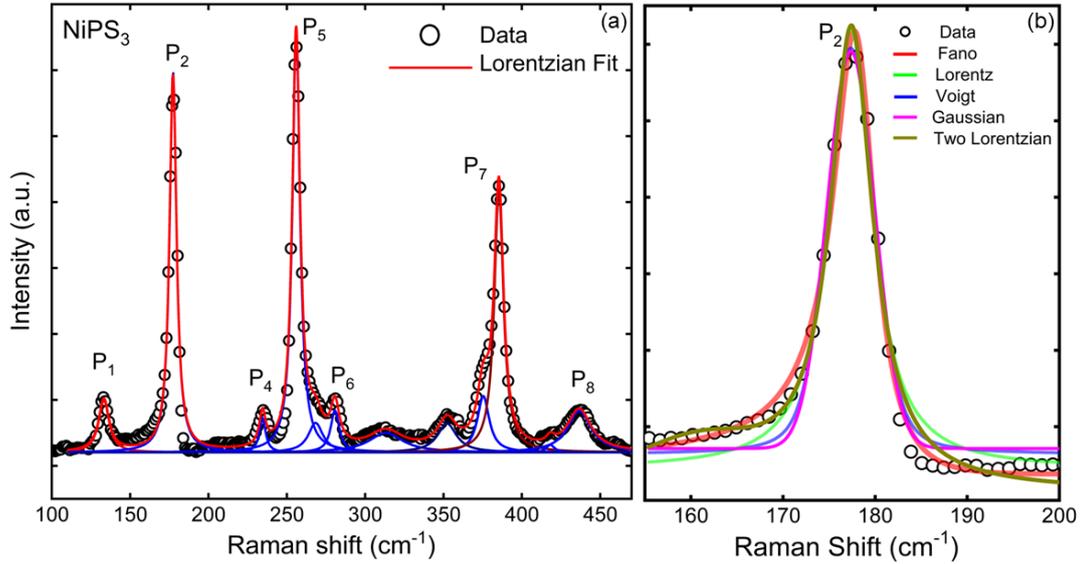

**Fig. 4**. (a) Unpolarized room-temperature Raman spectra of $NiPS_3$ single crystal. The open circle represents the experimental data and the red solid line is the Lorentzian fit of the whole spectra. (b) A zoomed-in view of the peak $P_2$. The best-fits corresponding to Fano, Lorentzian, Voigt, Gaussian, and two Lorentzian lineshapes are also shown.

$$I(\omega) = I_0 \frac{[1+2(\omega-\omega_0)/((q\Gamma))]^2}{1+4((\omega-\omega_0)^2)/\Gamma^2} \tag{1}$$

where, $I_0$ is the peak intensity, $1/q$ is the coupling coefficient which describes the coupling of the discrete phonon mode in question with an excitation continuum, $\Gamma$ is the FWHM, and $\omega_0$ is

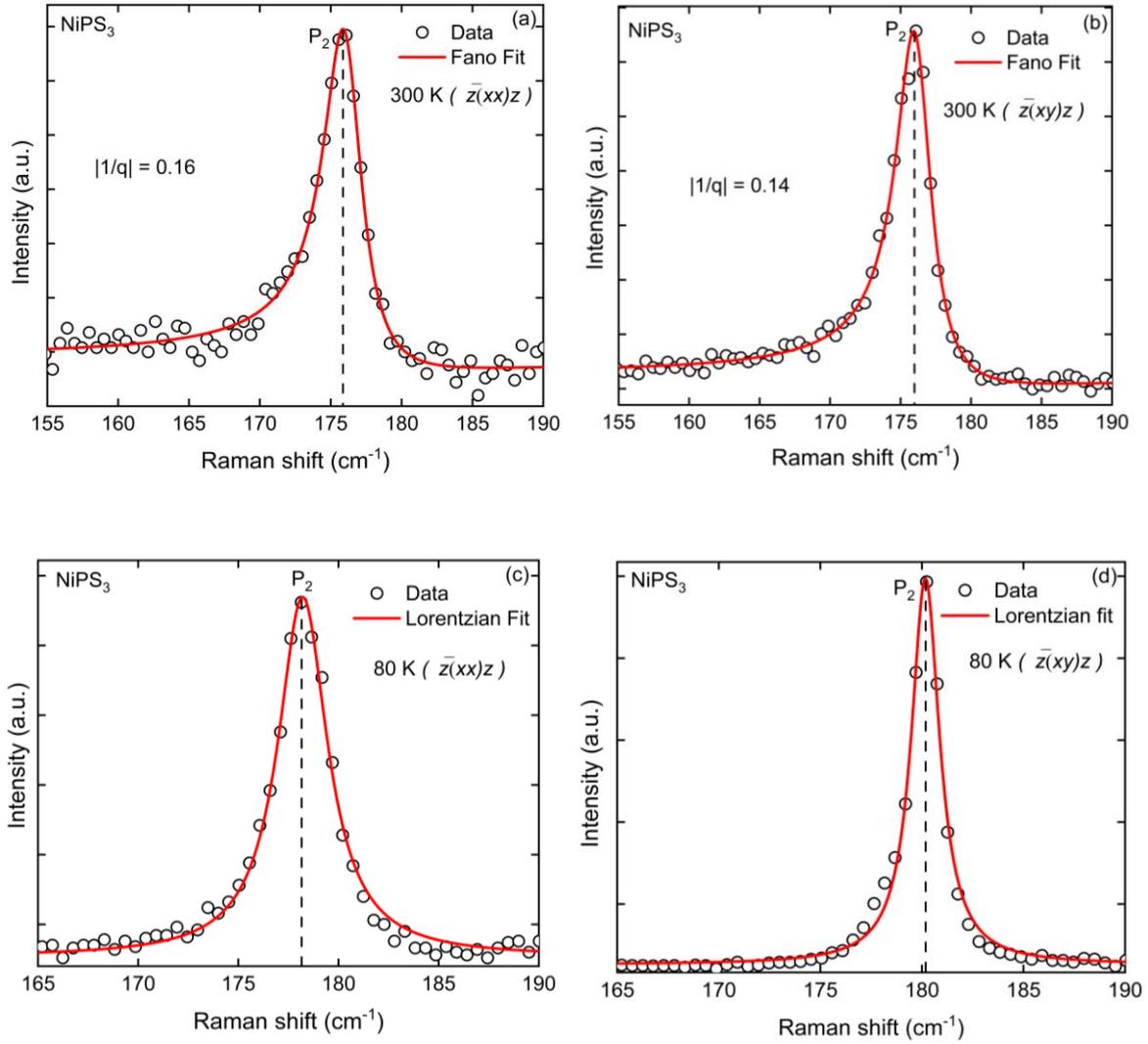

**Fig. 5**. The Raman mode P$_2$ in a NiPS$_3$ single crystal measured under two parallel [$\bar{z}xxz$] and cross [$\bar{z}xyz$] polarization configurations at 80 K and 300 K. (a) [$\bar{z}xxz$] at 300 K, (b) [$\bar{z}xyz$] at 300 K, (c) [$\bar{z}xxz$] at 80 K, and (d) [$\bar{z}xyz$] at 80 K.

the phonon frequency in the absence of the coupling. The fit using Fano lineshape is shown in Fig. 4b. The dimensionless parameter $q$ quantifies the degree of asymmetry in the lineshape. As $1/q$ approaches *zero*, the Fano lineshape becomes Lorentzian, marking the disappearance of Fano resonance. In NiPS$_3$, for the P$_2$ phonon, the $|1/q|$ value at room temperature is 0.13, which is comparable to previously reported values for other well investigated systems where Fano resonance has been reported, including (a) metallic carbon nanotubes ($|1/q|$: 0.04 to 0.10 depending on the diameter of the nanotube) [36], (b) LaMnO$_{3+\delta}$ thin films ($|1/q|$: 0.2 to

0.4 depending on temperature and oxygen partial pressure during growth) [37], (c) Monolayer graphene ($|1/q|$: 0.07) [38], (d) (Ca/Sr)Pd$_3$O$_4$ compounds ($|1/q|$: 0.07 to 0.2 depending on temperature) [39], and few-quintuple layers topological insulator Bi$_2$Se$_3$ nanoplatelets ($|1/q|$: 0.047 to 0.084) [40]. In Fig. 5, we show a zoomed in view of P$_2$ at 80 K and 300 K in the parallel [$\bar{z}xxz$] and cross [$\bar{z}xyz$] polarization. At 80 K, in both polarizations the lineshape is symmetric and can be fitted satisfactorily using the Lorentzian lineshape. However, at 300 K, a clear characteristic Fano asymmetry, absent at 80 K, can be seen for both polarization configurations. Moreover, at 80 K, the positions $P_2^{\parallel} = 178.1$ cm$^{-1}$ and $P_2^{\perp} = 180.2$ cm$^{-1}$, giving $\Delta P = 2.1$ cm$^{-1}$ in excellent agreement with Ref. [24]. Not only are two peaks occurring at two distinct positions, their FWHMs also differ substantially with $FWHM^{\parallel} = 3.3$ cm$^{-1}$, and $FWHM^{\perp} = 2.0$ cm$^{-1}$, giving $\Delta FWHM = 1.3$ cm$^{-1}$. On the other hand, at 300 K, the two configurations become degenerate with $P_2^{\parallel} = P_2^{\perp} = 176.0$ cm$^{-1}$ and $FWHM^{\parallel} = FWHM^{\perp} \cong 3.4$ cm$^{-1}$.

To probe this further, we measured the Raman spectra over a broad temperature range from 80 K to 400 K. Since in the cross $\bar{z}xyz$-polarization configuration the quasi-elastic scattering (QES) contribution is negligible [24], the Raman data in the $\bar{z}xyz$-polarization configuration for pristine NiPS$_3$ are shown in Fig. 6. At 80 K, the most prominent feature is the presence of a broad peak centered at ~550 cm$^{-1}$. This peak is due to the two-magnon scattering. The phonon P$_9$ near 570 cm$^{-1}$ exhibits a Fano lineshape below T$_N$ in agreement with Ref. [24]. The Fano resonance of the P$_9$ phonon arises from its coupling with the two-magnon continuum, which has been established convincingly in the previous work [24]. Here, we focus on the mode P$_2$ which we now know exhibits Fano line shape, but only at high temperatures. By fitting P$_2$ at various temperatures using the expression for the Fano lineshape in eq. (1), we plot the temperature variation of parameter $q$ in NiPS$_3$ in Fig. 7. The derivative $dq/dT$ is also shown for clarity. We see that upon cooling the sample below T$_N$, $q$ increases sharply, suggesting that the Fano resonance associated with $P_2$ vanishes at low-temperatures in the magnetically ordered phase. The corresponding data for $Ni_{0.92}Zn_{0.08}PS_3$ (x = 0.08), for which $T_N$ has reduce to 135 K (from 155 K for the pristine $NiPS_3$) are also shown in Fig. 7. In this case also a similar behavior is observed with Fano asymmetry disappearing below 135 K. This supports the observation that the Fano resonance of $P_2$ in $NiPS_3$ disappears gradually below $T_N$. This is contrary to the behavior of mode $P_9$ where Fano resonance appears below $T_N$ but disappears gradually at higher temperatures. Furthermore, we note the sign of parameter $q$ is *negative* for both $P_2$ and $P_9$; in other words, the asymmetric broadening of both cases is on the lower-wavenumber side of the peak. For $P_9$ this is consistent with the fact that that two-magnon excitation continuum has its peak at (550 $cm^{-1}$), which is on the lower-wavenumber side of the mode $P_9$ at 550 $cm^{-1}$ resulting in an antiresonance between the discrete phonon $P_9$ and the two-magnon continuum.

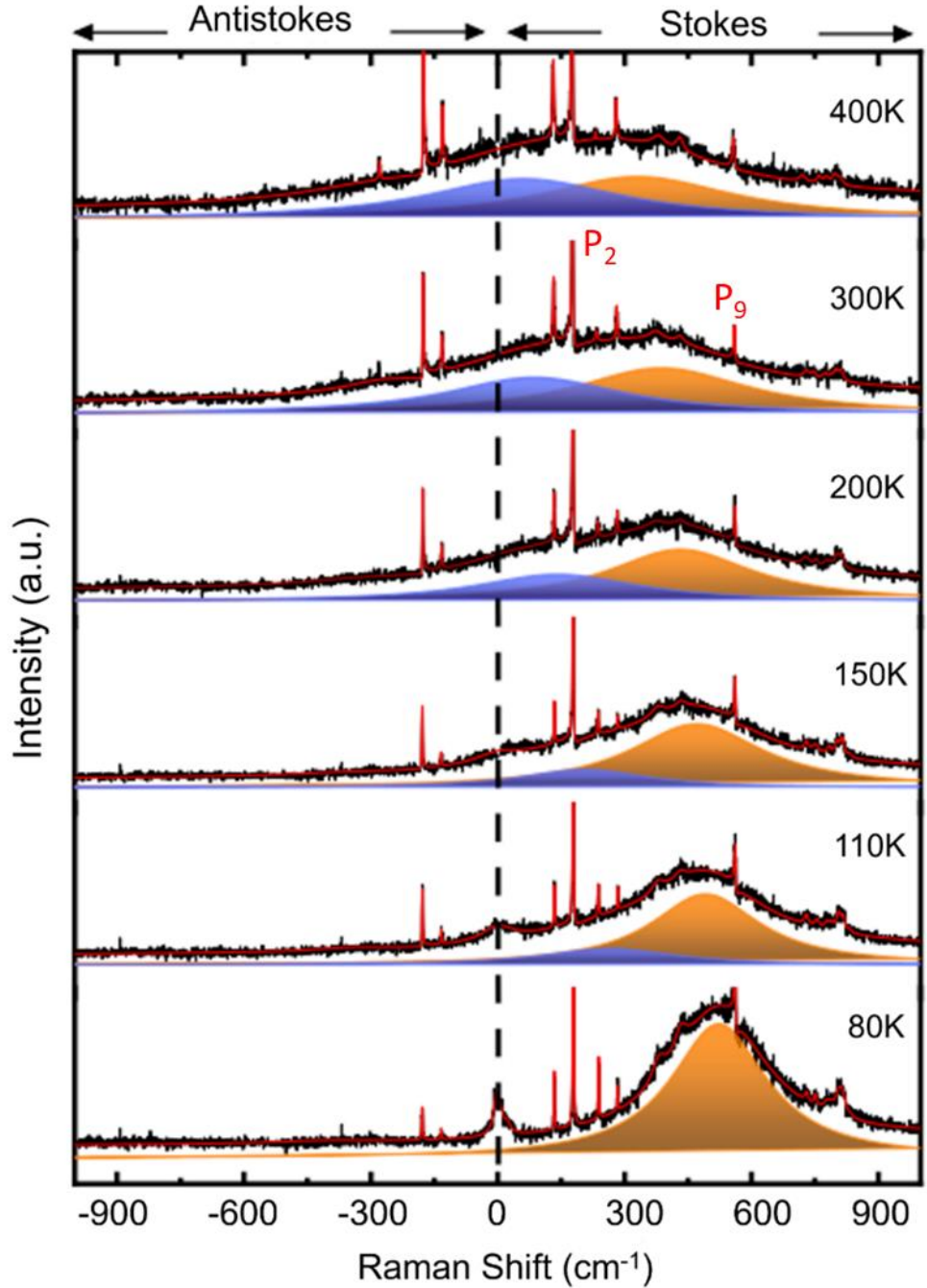

**Fig. 6.** Raman Spectra in the Stokes and Anti-Stokes region for NiPS$_3$ at various temperatures between 400 K and 80 K. The orange shaded region represents the two-magnon continuum and the blue shaded region represents the second continuum which appears at high temperatures. The red line through the data points is a fit to the data.

In order to understand the origin of Fano resonance of $P_2$ in the paramagnetic phase, we notice that the broad continuum in the Raman spectra of $NiPS_3$ above $T_N$ is best described by considering two broad peaks as shown in Fig. 6 using the orange and blue shades. The orange-shaded peak is the familiar two-magnon continuum which is the only continuum that remains at low temperatures (for example, at 80 K it peaks around 550 $cm^{-1}$ as shown previously in Ref. [24]). However, the spectra shown at 200 K and above necessarily require two broad peaks to be included in the fitting in order to get a satisfactory fit. As noted above, we reiterate that the quasi-elastic component is not significant in the cross-polarization at high temperatures for which the data shown. In fact, as shown in Fig. 6, at 80 K, a small QES signal can be seen in the form a peak around 0 $cm^{-1}$; however, the intensity of this signal suppresses rapidly to become insignificant upon warming. Thus, we can rule out QES as the origin of the second (blue shaded) continuum. We argue that the Fano lineshape of $P_2$ is consequence of its destructive interference (antiresonance) with this continuum. As this continuum peaks towards the lower-wavenumber side of P$_2$, the sign of Fano parameter $q$ should be negative, which is consistent with the sign of $q$ for P$_2$ observed experimentally. The negative sign of $q$ for both P$_2$ and P$_9$ is consistent with the observation that the peak in their respective continuums appear on the lower-wavenumber side of these modes, indicating antiresonance or destructive interference (see for example [41]).

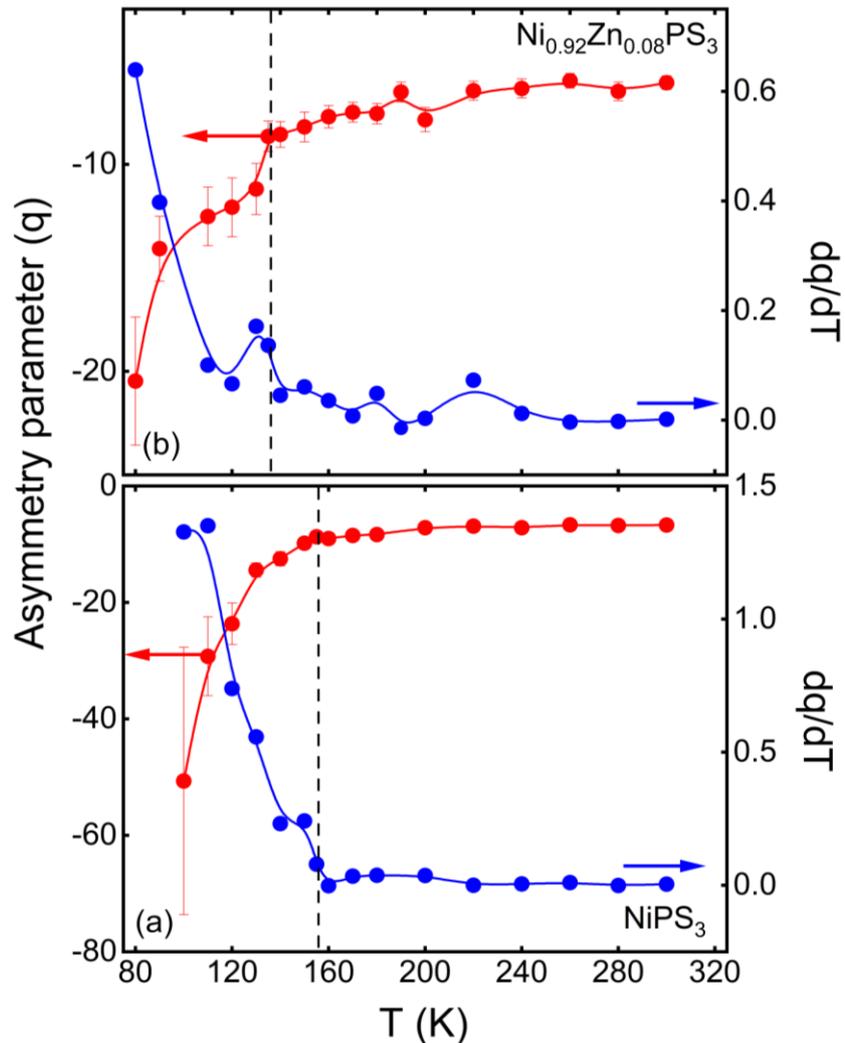

**Fig. 7**. The temperature variation of asymmetry parameter ($q$) in (a) NiPS$_3$ and (b) Ni$_{0.92}$Zn$_{0.08}$PS$_3$. The derivative $dq/dT$ is also shown. The dashed line indicates the antiferromagnetic ordering temperature (T$_N$).

We now discuss on the possible origin of the second continuum at high temperatures. From previous studies, we know that NiPS$_3$ is a 'self-doped' negative charge transfer (NCT) insulator [16]. Therefore, the Ni ion in NiPS$_3$ has a valence state that is a linear superposition of $d^8$, $d^9\underline{L}$, and $d^{10}\underline{L}^2$ configurations, where $\underline{L}$ refers to a ligand hole [16]. We speculate that the second broad continuum at low wavenumbers is likely associated with the continuum of charge states formed by $d^8$, $d^9\underline{L}$, and $d^{10}\underline{L}^2$ configurations. The discrete phonon mode $P_2$, which lies in the same energy range, then couples with the continuum of electronic states $\psi = \alpha d^8 + + \beta d^9\underline{L} + \gamma\, d^{10}\underline{L}^2$ to give rise to the Fano resonance. To show this convincingly one needs to establish the energy and symmetry equivalence of the continuum of electronic states $\psi$ and phonon $P_2$, which is beyond the scope of the present study. However, we indirectly corroborate this hypothesis using the experimental observations concerning the size of coupling constant $|1/q|$ in the Zn-doped samples in the next paragraph.

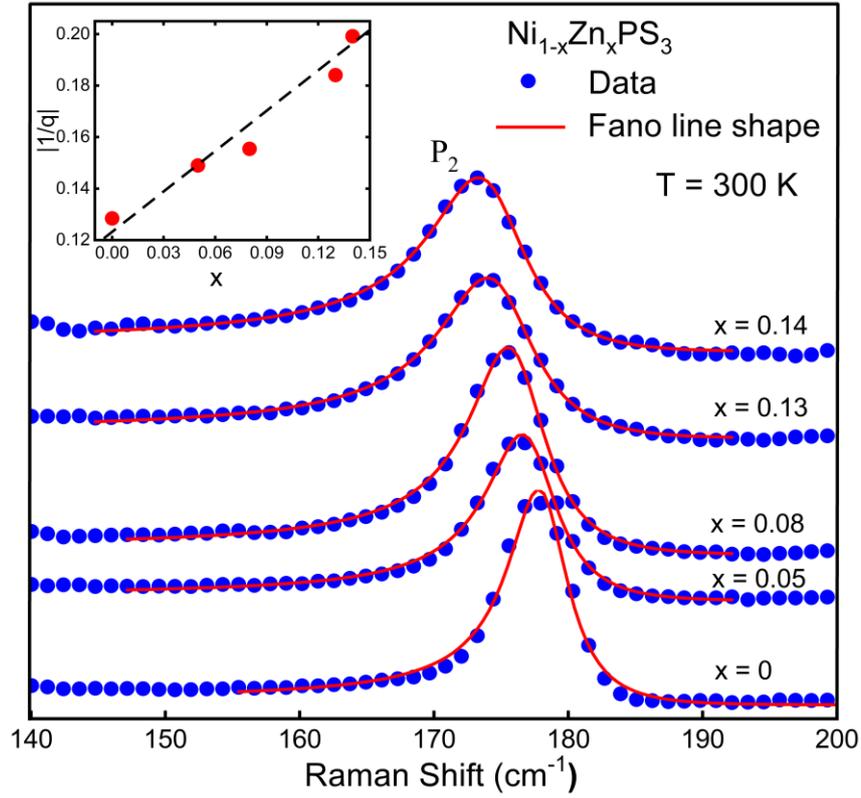

**Fig. 8.** The Raman mode $P_2$ in $Ni_{1-x}Zn_xPS_3$ crystals at 300 K for x = 0, 0.05, 0.08, 0.13, and 0.14. Inset shows the variation of coupling coefficient $|1/q|$ as a function of x.

The $P_2$ for all the different Zn doped samples at 300 K is shown in Fig. 8. With increasing Zn, the mode $P_2$ undergoes a redshift; its intensity shows a monotonic decrease while the

$FWHM$ shows an increasing trend. The observed redshift is due to the higher ionic mass of Zn compared to Ni, and the increasing $FWHM$ can be attributed to the disorder due to Zn-doping. The Fano lineshape fits for different Zn concentrations is shown in Fig. 8. With increase in Zn doping, the coefficient $|1/q|$, quantifying the peak asymmetry, increases monotonically with the doping concentration as shown in the inset of Fig. 8. Since the laser wavelength in our experiments is fixed, the increase in the asymmetry cannot be due to the laser intensity [42].

To explain this increasing trend, we propose a mechanism, shown as a cartoon in Fig. 9. In the crystal structure of NiPS$_3$, each sulfur is bonded to two neighboring Ni ions as shown in Fig. 9(a). The negative charge transfer from S to Ni therefore takes place over two possible pathways that are shown using the black arrows in Fig. 9(a). On the other hand, when Zn occupies one of the two Ni sites, the average number of Ni ions bonded to the central S will be less than 2. As the d orbitals of Zn are fully occupied ($d^{10}$), the bridging S has now only one Ni site available to it to transfer the same quantity of electronic charge as before, as shown in Fig. 9(b). Thus, in the presence of Zn doping, the average negative charge transferred to Ni should be more, with the result that the weights of the $d^9\underline{L}$ and $d^{10}\underline{L}^2$ components should enhance, which, in turn, should enhance the charge-phonon coupling further. To check the validity of this scenario (that the weights of $d^9\underline{L}/d^{10}\underline{L}^2$ component enhance), we carried out XPS on a NiPS$_3$ and Ni$_{0.92}$Zn$_{0.08}$PS$_3$. The $2p$ core-level XPS spectra for the two samples are shown in Fig. 10. The XPS spectra of NiPS$_3$ consists of three-peaks positioned at binding energies (BE) 855 eV (M$_1$), 860 eV (M$_2$) and 865.6 eV (M$_3$). The presence of these three peaks is a signature of NCT state with $d^8$, $d^9\underline{L}$, and $d^{10}\underline{L}^2$ components, as shown theoretically as well as experimentally in the well-established NCT compound NiGa$_2$S$_4$ [17]. Our data are also in excellent agreement with previously reported XPS for NiPS$_3$ [16]. Our main finding is that for Ni$_{0.92}$Zn$_{0.08}$PS$_3$ the spectrum shifts to lower binding energies, which indicates that the weights of $d^9\underline{L}$ and $d^{10}\underline{L}^2$ components has increased due to Zn doping. This suggests a strong charge-phonon coupling in the paramagnetic state of NiPS$_3$. Why the spin-phonon coupling appears in the magnetically ordered state is easy to understand, but what weakens the charge-phonon coupling at low temperatures is not very clear. One can speculate that the antiferromagnetic ordering of the Ni spins results in renormalization of the electronic structure [43]. The other possibility is that at low-temperatures the charge density fluctuations, i.e., electron delocalization between $d^8$, $d^9\underline{L}$, and $d^{10}\underline{L}^2$ components, freeze out.

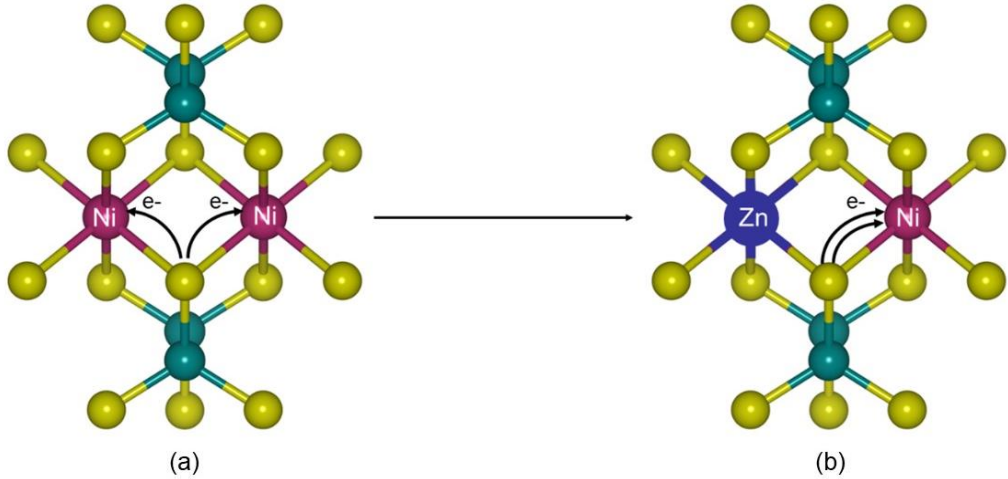

**Fig. 9.** The Schematic depiction of charge transfer from Ni to S in the negative charge transfer insulator NiPS$_3$. In the left structure, each S atom is transferring electrons to two Ni atoms, upon Zn doping as shown in the right structure each S atom is now transferring electrons to only one Ni atom and hence electron-phonon coupling is increased with Zn doping.

## II. Summary and conclusions

We have grown a high-quality of single crystals of Ni$_{1-x}$Zn$_x$PS$_3$ (x = 0, 0.01, 0.03, 0.05, 0.08, 0.13, 0.14) using physical vapor transport technique and characterized them using x-ray diffraction, FESEM, EDX and HRTEM probes. The zigzag antiferromagnetic ordering temperature ($T_N$) is found to suppress with increasing Zn doping. The rate of suppression $dT_N/dx$ follows the mean-field prediction for spin 1 on a honeycomb lattice [29]. In the previous Raman, Photoluminescence and XPS studies, NiPS$_3$ has been shown to exhibit several interesting features, indicative of the presence of spin-phonon and spin-charge coupling [24,31–33,44,45].

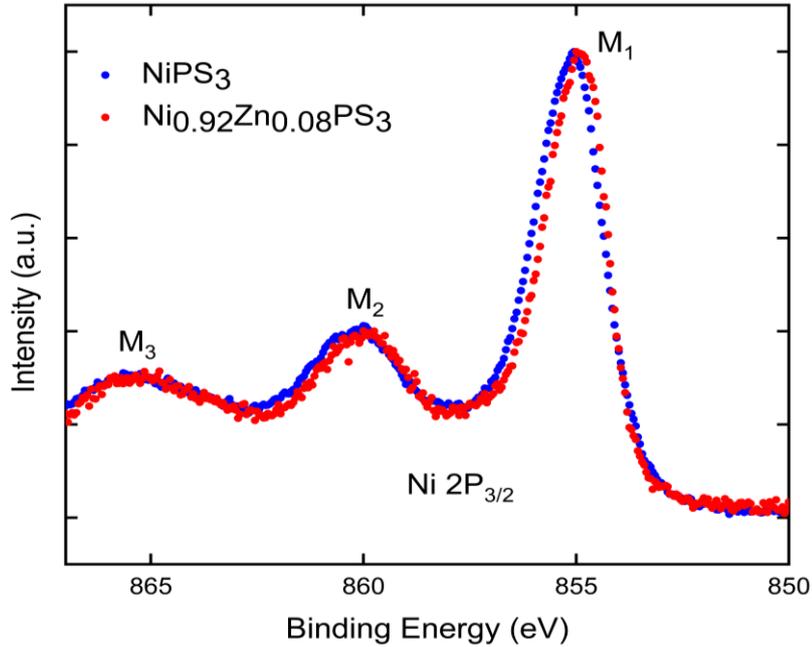

**Fig. 10.** 2p core level XPS spectra of NiPS$_3$ (Blue circles) and Ni$_{0.92}$Zn$_{0.02}$PS$_3$ (Red circles).

Here, we show that the Raman mode due to the vibration of Ni near 176 cm$^{-1}$ (called P$_2$) exhibits a Fano-type asymmetry at high temperatures ($T > T_N$). The measurements carried out, both, below and above $T_N$, and under two different polarization configurations, confirm that P$_2$ exhibits a Fano resonance above $T_N$. We show that the sign of Fano asymmetry parameter $q$ for both P$_2$ and P$_9$ (near 550 cm$^{-1}$) is negative. We argue that the Fano resonance of P$_2$ arises from its coupling with a broad electronic continuum that is centered slightly towards the lower-wavenumber side of P$_2$. This electronic continuum suppresses rapidly below $T_N$ and at the same time the Fano-like asymmetry of P$_2$ also disappears, clearly indicating a correlation between the two. This correlation was further supported by a similar observation in the temperature dependent experiments on a Zn-doped sample with reduced $T_N$. We further show that Zn doping enhances the Fano asymmetry. As NiPS$_3$ is a negative charge transfer (NCT) insulator, with the electronic states of Ni describes as $\psi = \alpha d^8 + + \beta d^9 \underline{L} + \gamma d^{10} \underline{L}^2$, we argue that the Fano resonance occurs due to the coupling of discrete phonon P$_2$ with the continuum of electronic states $\psi$. If we assume that upon small Zn doping the amount of charge transfer from S ($p$) to Ni ($d$) orbitals remains unaffected, then the amount of charge transferred to 1 Ni atom per formula unit in NiPS$_3$ is now transferred to (1-x) Ni atoms in Ni$_{1-x}$Zn$_x$PS$_3$, which implies that the weights of the $d^9\underline{L}$ and $d^{10}\underline{L}^2$ components enhance, which, in turn, enhances the charge-phonon coupling. The peaks in the $2p$ core-level XPS of Ni in NiPS$_3$ are shown to redshift upon Zn-doping supporting the argument that the weights of $d^9\underline{L}/d^{10}\underline{L}^2$ components increase upon Zn-doping. To summarize, we show that NiPS$_3$ is a unique 2D Van der Waals magnetic insulator, which exhibits charge-phonon coupling in its paramagnetic state and spin-phonon couplings in the antiferromagnetically ordered phase.

**Acknowledgements:**

NP would like to thank CSIR for the PhD fellowship. AK acknowledge PMRF PhD fellowship. SS and LH would like to thank QTF I-HUB, IISER Pune, India for supporting research on 2D quantum materials.

**Appendix:**

The single crystals of $Ni_{1-x}Zn_xPS_3$ were grown using the physical vapor transport technique whose details are given in the main text. The grown crystals exhibit layered morphology. A few representative images of the single crystals and their corresponding In-lens images are shown in the main text. The actual or measured composition x of Zn in $Ni_{1-x}Zn_xPS_3$ crystals is obtained using EDX. The nominal and EDX composition of the grown crystals are summarized in Table 1 below. The homogeneity of the grown crystals was further verified through chemical mapping as shown in Fig. A1. The Phase purity of the grown crystals was verified using Powder X-Ray diffraction. Fig. A2 (a) shows the Powder x-ray diffraction pattern and the Fig. A2 (b) shows the monotonic increase in the lattice parameter due to higher ionic radii of $Zn^{2+}$ with respect to $Ni^{2+}$.

The 2p core level XPS spectra for the elements P and S in $NiPS_3$ and $Ni_{0.92}Zn_{0.08}PS_3$ is shown in Fig. A3 and Fig. A4, respectively. As shown in Fig. A3, the $2p$ spin-orbit split doublet for P in $NiPS_3$ appear in the binding energy range 131 eV to 135 eV. In this range, besides the spin-orbit split doublet, no extra peak can be seen. On the other hand, in the Zn-doped sample the spectra could only be fitted satisfactorily when two doublets are considered. This difference can be attributed to the presence of two different chemical environments in the doped sample. Similarly, the S $2p$ spectra can be fitted to one doublet as expected in the pristine sample, and a second spin-orbit doublet had to be considered to account for the whole spectra in the presence of Zn-doping as shown on Fig. A4.

Table 1. The Nominal and EDX composition of the grown $Ni_{1-x}Zn_xPS_3$ single crystals

| Nominal composition | EDX composition |
|---|---|
| $NiPS_3$ | $NiPS_3$ |
| $Ni_{0.99}Zn_{0.01}PS_3$ | $Ni_{0.99}Zn_{0.01}PS_3$ |
| $Ni_{0.98}Zn_{0.02}PS_3$ | $Ni_{0.97}Zn_{0.03}PS_3$ |
| $Ni_{0.97}Zn_{0.03}PS_3$ | $Ni_{0.95}Zn_{0.050}PS_3$ |
| $Ni_{0.90}Zn_{0.10}PS_3$ | $Ni_{0.92}Zn_{0.08}PS_3$ |
| $Ni_{0.85}Zn_{0.15}PS_3$ | $Ni_{0.87}Zn_{0.13}PS_3$ |
| $Ni_{0.80}Zn_{0.20}PS_3$ | $Ni_{0.86}Zn_{0.14}PS_3$ |

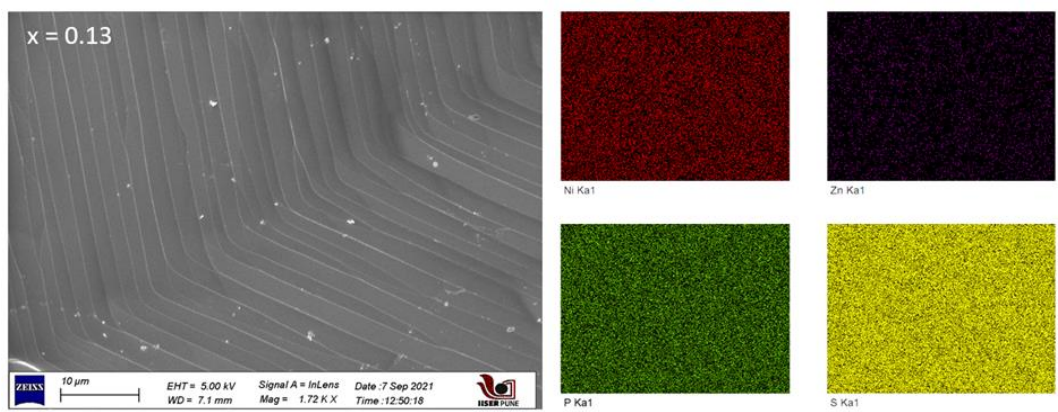

**Fig. A1.** FESEM-images of single crystals of $Ni_{1-x}Zn_xPS_3$ (x = 0.13) investigated using in - Lense detector exhibiting layered structure

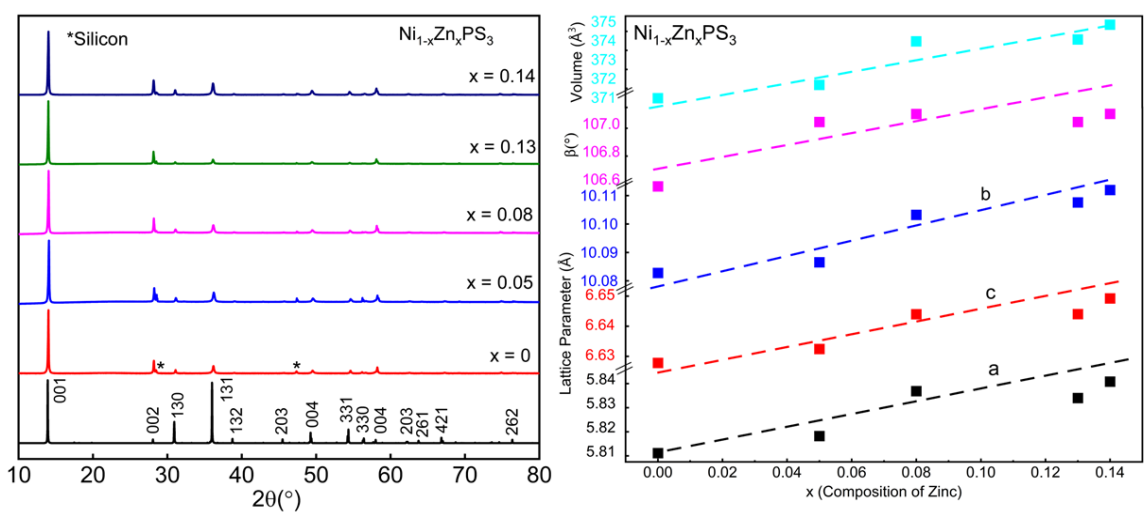

**Fig. A2.** (a) Powder X-ray diffraction pattern for different Zn composition using Si as a standard to estimate accurate lattice parameter. (b) Trend in lattice parameter with x (Zn composition).

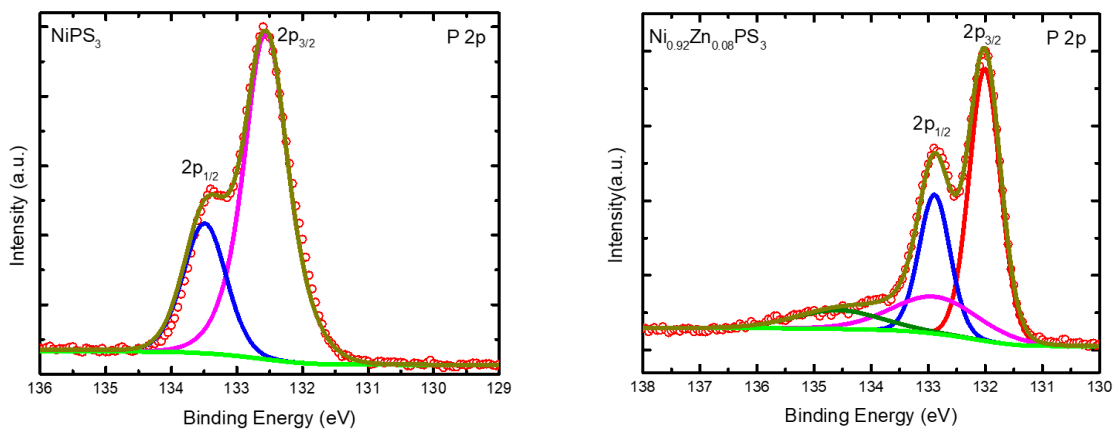

**Fig. A3.** The red open circles, the dark yellow solid line and the green solid line indicates the experimental data, envelope of all the fitted peaks and the Shirley background respectively. (a) 2p core level spectra of P in NiPS$_3$. (b) 2p core level spectra of P in Ni$_{0.92}$Zn$_{0.08}$PS$_3$

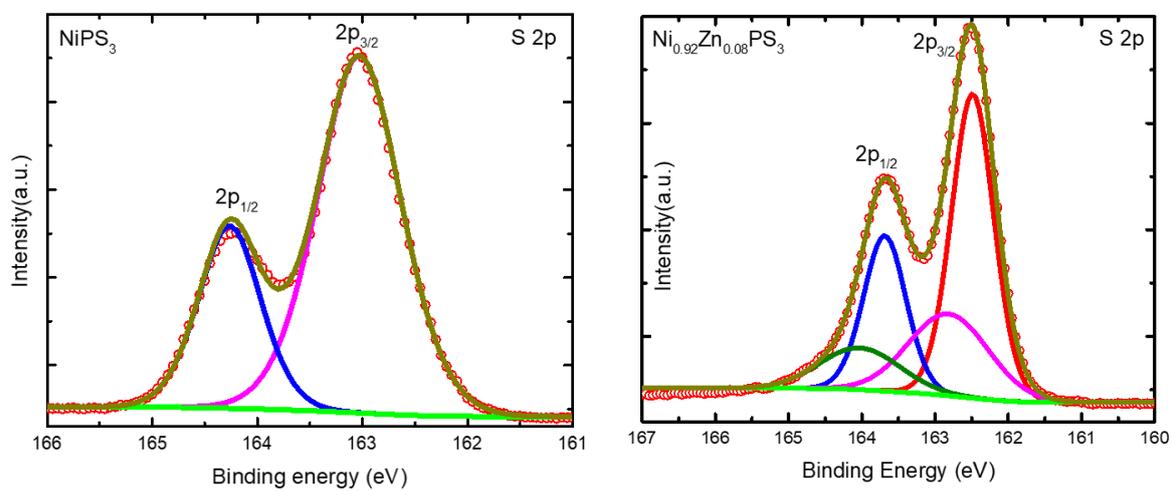

**Fig. A4.** The red open circles, the dark yellow solid line and the green solid line indicates the experimental data, envelope of all the fitted peaks and the Shirley background

respectively. (a) 2p core level spectra of S in NiPS$_3$. (b) 2p core level spectra of S in Ni$_{0.92}$Zn$_{0.08}$PS$_3$